\documentclass[reprint, superscriptaddress,amsmath,amssymb,aps,prl]{revtex4-2}
\usepackage{hyperref}
\hypersetup{colorlinks=true, citecolor=blue, urlcolor=blue, linkcolor=blue}
\usepackage{graphicx, nicefrac}
\usepackage{xcolor}
\usepackage{physics}
\usepackage{upgreek}
\usepackage{array}

\usepackage{multirow}

\usepackage{verbatim}

\usepackage{braket}

\usepackage{xspace} 
\newcommand{\ba}{$^{137}$Ba$^+$\xspace}

\makeatletter
\newcommand{\thickhline}{%
    \noalign {\ifnum 0=`}\fi \hrule height 1pt
    \futurelet \reserved@a \@xhline
}
\newcolumntype{"}{@{\hspace{1pt}\hskip\tabcolsep\vrule width 1pt\hskip\tabcolsep\hspace{1pt}}}
\makeatother

\begin{document}
\title{High fidelity state preparation and measurement of ion hyperfine qubits with $I>\frac{1}{2}$}
\author{Fangzhao Alex An}
\email{Fangzhao.An@Quantinuum.com}
\affiliation{Quantinuum, Golden Valley, MN, USA}
\author{Anthony Ransford}
\email{Anthony.Ransford@Quantinuum.com}
\affiliation{Quantinuum, Broomfield, CO, USA}
\author{Andrew Schaffer}
\author{Lucas R. Sletten}
\affiliation{Quantinuum, Golden Valley, MN, USA}
\author{John Gaebler}
\affiliation{Quantinuum, Broomfield, CO, USA}
\author{James Hostetter}
\author{Grahame Vittorini}
\affiliation{Quantinuum, Golden Valley, MN, USA}

\date{\today}

\begin{abstract}
We present a method for achieving high fidelity state preparation and measurement (SPAM) using trapped ion hyperfine qubits with nuclear spins higher than $I = 1/2$.
The ground states of these higher nuclear spin isotopes do not afford a simple frequency-selective state preparation scheme.
%
We circumvent this limitation by stroboscopically driving strong and weak transitions, blending fast optical pumping using dipole transitions and narrow microwave or optical quadrupole transitions.
%
%
We demonstrate this method with the $I=3/2$ isotope $^{137}\mbox{Ba}^+$ to achieve a SPAM infidelity of $\left(9.0 \pm 1.3\right) \times 10^{-5}$ ($-40.5 \pm 0.6$ dB), facilitating the use of a wider range of ion isotopes with favorable wavelengths and masses for quantum computation.
\end{abstract}

\maketitle


State preparation and measurement (SPAM) is fundamental to quantum computation and covers two of the five necessary DiVincenzo criteria~\cite{divincenzo2000physical}.
Modern quantum systems are not error corrected, and the fidelity of an uncorrected $N$-qubit register typically decreases exponentially with size as $(\mathcal{F}_\mathrm{SPAM})^N$, where $\mathcal{F}_\mathrm{SPAM}$ is the single-qubit SPAM fidelity.
Current error correction codes require mid-circuit measurement and reset (MCMR) and thus SPAM errors contribute to the error correction budget as the ratio of gates to measurements~\cite{ryan2021realization}.
Given the historic difficulty of improving gate fidelity, it will likely be desirable to make the SPAM contribution as small as possible.
%
Moreover, the speed at which SPAM can be performed can affect memory errors during MCMR, and the compatibility of the exact SPAM technique with other qubit operations must also be considered.
More broadly, it is desirable to have fast, high-fidelity SPAM in ionic qubits with other practical advantages to quantum computation such as light mass and visible wavelengths for gates.


Although the $^{171}\mbox{Yb}^+$ ion has a hyperfine clock qubit and its $I=1/2$ nuclear spin results in a singlet state for simple, high fidelity state preparation~\cite{olmschenk2007ybspam,crain2019ybspam}, the limitations on measurement fidelity, necessity for high power UV light, and associated photo-induced charging~\cite{wang2011charging} present significant engineering challenges.
Other systems with similar structure (nuclear spin $I=1/2$) have been explored, with the most promising being $^{133}\mbox{Ba}^+$~\cite{Ba133,lee2006quantum} due to its visible wavelength transitions, but these species either require lasers deep in the UV or are not naturally occurring, presenting formidable challenges for scaling systems.

\begin{figure}[!t]
		\centering
		\includegraphics[width=\linewidth]{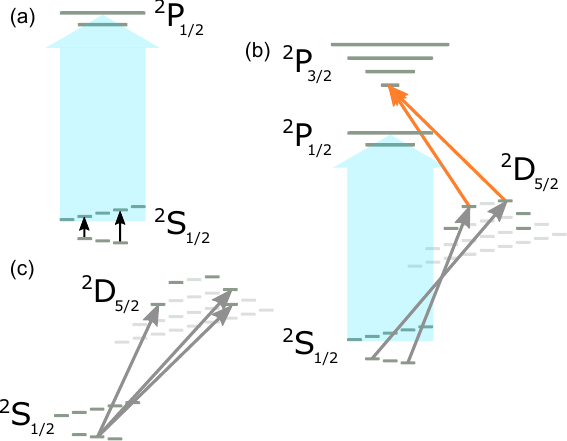}
		\caption{\label{Level Diagram}
			\textbf{(a)} Microwave-assisted optical pumping (MAOP) using coherent microwave pulses (black) and 493~nm flush pulses (aqua).
			\textbf{(b)} 1762~nm narrow-band optical pumping (NBOP) using coherent 1762~nm pulses (gray), 614~nm quenching (orange), and 493~nm flush (aqua).
			\textbf{(c)} Cabinet shelving with repeated 1762~nm pulses addressing $|0\rangle$. Hyperfine quantum numbers $F$ and $m_F$ are omitted for clarity (see Supplementary Materials for detailed labeling and timing diagram~\cite{SuppMats}).}
		\label{fig:level_diagram}
\end{figure}

Despite their complicated structure, ions with nuclear spin $I>1/2$ have advantages~\cite{low2020highspin,omg} and are used in quantum information and quantum computing, including $^{9}\mbox{Be}^+$~\cite{gaebler2016high}, $^{43}\mbox{Ca}^+$~\cite{benhelm2008ca,ballance2016high}, $^{25}\mbox{Mg}^+$~\cite{tan2015multi} and $^{137}\mbox{Ba}^+$~\cite{dietrich2010hyperfine,bramman2019thesis}.
Currently, all methods of state preparation in these species are limited in performance by their requirement for either light with high polarization purity or a series of coherent gate operations to map a prepared state to the clock qubit state, where this transfer infidelity shows up as a preparation error~\cite{harty2014high,gaebler2016high}.

\begin{table*}[!htbp]
    \centering
    \begin{tabular}{c|c|c|c|c|c|c|c}
    \hline \hline
        Species & $I$ & $\eta_{_{F' + 1, F + 1}}$ & $\eta_{_{F', F + 1}}$ & $\frac{\Gamma}{2\pi} (\mathrm{MHz})$ &Hyperfine S (GHz) &Hyperfine P (GHz)&$\epsilon_{\mathrm{prep}}$ \\
        \hline
         ${}^{9}\text{Be}^+$ & 3/2 & 1/2 & 5/6 & 22.4~\cite{langer2006high} &1.25~\cite{wineland1983laser}&0.194~\cite{bollinger1985hyperfine}& 5.4e-4  \\
         ${}^{25}\text{Mg}^+$ & 5/2 & 4/9 &7/9  & 42.4~\cite{nguyen2009linewidth} &1.788~\cite{nguyen2009linewidth}&0.307~\cite{nguyen2009linewidth}& 1.0e-3\\
         ${}^{43}\text{Ca}^+$ & 7/2 & 5/12& 3/4 & 22.4~\cite{james1997quantum} &3.226~\cite{lucas2004isotope}&0.581~\cite{lucas2004isotope}& 8.9e-5\\
         ${}^{87}\text{Sr}^+$ & 9/2 & 2/5 & 11/15 & 21.5~\cite{pinnington1995studies}&5.0~\cite{yuan1995hyperfine}&0.89~\cite{yuan1995hyperfine}& 3.4e-5 \\
         ${}^{135}\text{Ba}^+$ & 3/2 & 1/2 & 5/6& 20.1~\cite{christensen2020thesis} &7.18~\cite{villemoes1993ba135}&1.33~\cite{villemoes1993ba135}& 1.4e-5\\
         ${}^{137}\text{Ba}^+$ & 3/2 & 1/2 & 5/6& 20.1~\cite{christensen2020thesis} &8.03~\cite{dietrich2010hyperfine}&1.49~\cite{dietrich2010hyperfine}& 1.1e-5\\
         ${}^{173}\text{Yb}^+$ & 5/2 & 4/9 &7/9  & 19.7~\cite{olmschenk2009yb173} &10.5~\cite{munch1987yb173s}&1.85~\cite{roman2021thesis}& 6.3e-6 \\
         \hline \hline
    \end{tabular}
    \caption{Predicted state preparation infidelities in commonly used ion species. The infidelity generally decreases with mass with the exception of $^{25}$Mg$^+$, which has a transition linewidth that is approximately two times broader than the other transitions considered here.}
    \label{table:IonSpecies}
\end{table*}

In this work, we prepare a single $^{137}\mbox{Ba}^+$ ion into the qubit states $|0\rangle \equiv \Ket{6\text{S}_{1/2},F=1,m_F=0}$ and $|1\rangle \equiv \Ket{6\text{S}_{1/2},F=2,m_F=0}$ without the need of highly polarized light or coherent mapping of the qubit to the clock state.
By alternating between standard optical pumping on a dipole transition and driving narrow, state-selective microwave or optical quadrupole transitions, we are able to address and minimize the state preparation error in the Zeeman sublevels of the ground state $6\text{S}_{1/2}$.
%
This technique is both generalizable to many $I>1/2$ species and high performance, and we use it to achieve the highest SPAM fidelity recorded with any qubit to the best of our knowledge (previously $^{171}\mbox{Yb}^+$~\cite{ransford2021weak}).

We present two varieties of the state preparation scheme, both of which are cyclic in nature and seek to reduce state preparation errors in all Zeeman sublevels of $\text{S}_{1/2}$ except $\ket{0}$.
The microwave scheme, shown in Fig.~\ref{fig:level_diagram}(a), is similar to the microwave-assisted optical pumping (MAOP) recently performed on an atomic ensemble~\cite{tretiakov2021engineering}.
We first use dipole optical pumping at $493$~nm to ``flush'' out errors from the $(\text{S}_{1/2},F=2)$ manifold, redistributing the errors in the Zeeman sublevels of $(\text{S}_{1/2},F=1)$.
Next, errors in the non-qubit Zeeman sublevels of the $(\text{S}_{1/2}, F=1)$ manifold are moved up to $F=2$ with microwave $\pi$-pulses near the hyperfine splitting, and are subsequently flushed by the $493$~nm light when the cycle repeats.
In an alternate scheme, the errors in $(\text{S}_{1/2},F=1)$ can be addressed with narrow-band optical pumping (NBOP)~\cite{ransford2021weak} to $\text{D}_{5/2}$ on the 1762~nm quadrupole transition, and then redistributed back into the $(\text{S}_{1/2},F=1)$ manifold with a $614$~nm pulse (Fig.~\ref{fig:level_diagram}(b)).

State preparation with either MAOP or NBOP can be generalized to different ion species with high nuclear spin, and we model MAOP state preparation for a general alkaline earth ion with a dipole transition from S (hyperfine states $F, F+1$) to P ($F',F'+1$)~\cite{SuppMats}.
In the limit of many cycles and low flush beam power, we derive the state preparation error to be
\begin{equation}
\begin{aligned}
\epsilon_\mathrm{prep} = \frac{\Gamma^2}{2}\Bigg[\frac{1}{\delta_{_{\mathrm{HF}, \mathrm{S}}}^2} + \frac{\eta_{_{F', F + 1}}}{\eta_{_{F' + 1, F + 1}}}\frac{1}{\delta_{-}^2}\Bigg],
\label{eq:eprep}
\end{aligned}
\end{equation}
\noindent where $\Gamma$ is the transition linewidth, $\delta_{\mathrm{HF},\mathrm{S}}$ is the ground state hyperfine splitting, and $\eta_{i, j}$ denotes the branching from state $i$ to state $j$.
The first term is identical to twice the limit of state preparation in an $I=1/2$ ion and the second term is a correction factor that depends on the branching ratio and hyperfine splitting difference between the ground and excited states $\delta_{-}$.
The parameters for Eq.~\eqref{eq:eprep} and the associated fidelities for various $I>1/2$ hyperfine qubits are shown in Table~\ref{table:IonSpecies}.

\begin{figure*}[!t]
\centering
\includegraphics[width=\textwidth]{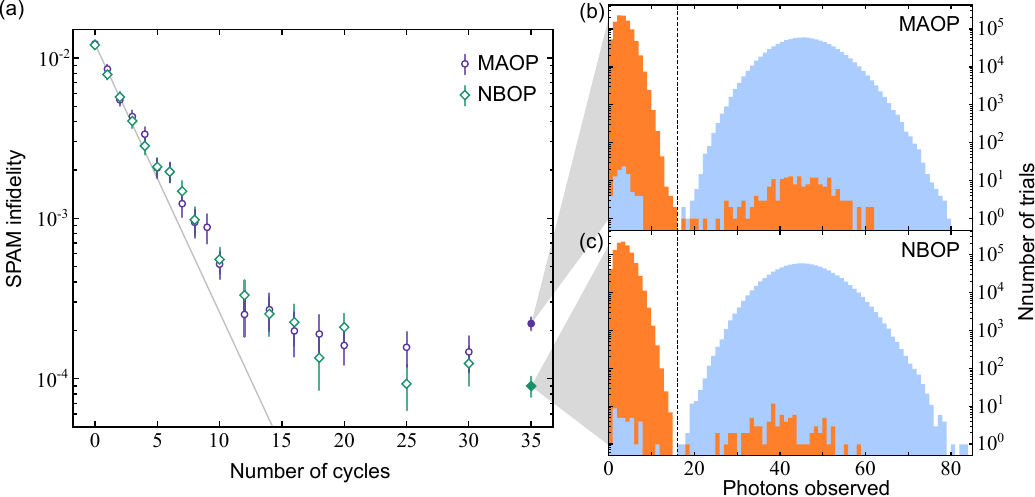}
\caption{\textbf{(a)} SPAM infidelity vs. increasing cycles of microwave-assisted (purple, MAOP) and narrow-band (green, NBOP) optical pumping. Infidelities are calculated from datasets with $50\times10^3$ (for 0 to 9 cycles), $100\times10^3$ (10 to 18 cycles), and $200\times10^3$ (20 to 30 cycles) trials. Both techniques are initially close to 1/3 reduction in error per cycle (gray line), but level off at high fidelity.
Solid points and fluorescence count histograms represent separate datasets for \textbf{(b)} MAOP (35 cycles) and \textbf{(c)} 1762~nm NBOP (30 cycles with 493~nm flush pulse and 5 cycles without flush), preparing the $\ket{0}$ (orange) and $\ket{1}$ (blue) qubit state over $10^6$ trials per state.
The SPAM infidelities were measured to be $(14.4 \pm 1.7)\times10^{-5}$ (microwaves) and $(6.1 \pm 1.1)\times10^{-5}$ (1762~nm) ($(22.0 \pm 2.1)\times10^{-5}$ and $(9.0 \pm 1.3)\times10^{-5}$ including leakage errors, respectively) using state detection thresholds indicated by vertical lines.
All error bars denote one Wilson interval.
}
\label{fig:histograms}
\end{figure*}

There are a variety of high fidelity measurement schemes in general, from quantum logic and quantum non-demolition~\cite{erickson2021high} to coherent and incoherent shelving.
For $^{137}\mathrm{Ba}^+$ we use a ``cabinet shelving'' procedure~\cite{benhelm2008ca} shown in Fig.~\ref{fig:level_diagram}(c) where $\ket{0}$ is shelved three times in succession to multiple Zeeman sublevels in the metastable $\text{D}_{5/2}$ state ($30$~s lifetime~\cite{DstateLifetime}).
We then we apply resonant 493~nm light and repumping 650~nm light for $350~\mu$s to detect fluorescence.
The resulting bright and dark state histograms are well separated and the SPAM infidelity for 1762~nm NBOP is measured to be $(9.0\pm1.3)\times 10^{-5}$ with a simple discriminator (Fig.~\ref{fig:histograms}(c)).
The data is taken without any pre- or post-processing, such as discarding trials based on low fluorescence counts during Doppler cooling.
We do not perform statistical detection of D-state decays for our SPAM fidelity~\cite{langer2006high}, but consider the effect in the error budget.

In this experiment, we trap single $^{137}\mathrm{Ba}^+$ ions 70 $\mu$m above a planar surface trap similar to that described in Ref.~\cite{pino2020demonstration}, and apply a magnetic field of $4.96$~G to define the quantization axis.
Single qubit rotations between the states $\ket{0}$ and $\ket{1}$ are driven by a microwave horn tuned near the transition frequency of $8.038$~GHz.
The ion is Doppler cooled by cycling through the $\text{S}_{1/2} \leftrightarrow \text{P}_{1/2}$ (493~nm cycling) and $\text{D}_{3/2} \leftrightarrow \text{P}_{1/2}$ (650~nm repump) transitions.
To address the many hyperfine levels of \ba, frequency sidebands are applied to all of our lasers with electro-optic modulators (EOMs)~\cite{SuppMats}.
The 493~nm cycling light is centered on the $F=2 \leftrightarrow F=2$ ($2\leftrightarrow 2$) hyperfine transition, and an EOM applies sidebands to address the other possible hyperfine transitions ($1\leftrightarrow 1$, $1\leftrightarrow 2$, and $2\leftrightarrow 1$).
For state preparation and measurement, we address the narrow shelving transition $\text{S}_{1/2} \leftrightarrow \text{D}_{5/2}$ with a $1762$~nm laser locked to a high-finesse optical cavity.
This laser is red-detuned by $88$~MHz from the $\ket{\text{S}_{1/2}, 1,0} \rightarrow \ket{\text{D}_{5/2}, 3,2}$ transition to avoid off-resonant shelving, and different shelving transitions are addressed with sidebands generated by an EOM.
A $614$~nm laser is used to deshelve ions from $\text{D}_{5/2}$ to $\text{P}_{3/2}$, from which ions preferentially fall into $\text{S}_{1/2}$.

We begin our state preparation with a fast ($40~\mu$s) polarization-limited step.
We address the $(\text{S}_{1/2}, F=2) \rightarrow (\text{P}_{1/2}, F=2)$ transition with $\pi$-polarized $493$~nm light and stroboscopically apply sidebands to address the $2 \leftrightarrow 1$ and $1 \leftrightarrow 1$ transitions (see Supplementary Materials for exact pulse sequences~\cite{SuppMats}).
Because of the forbidden selection rule $\ket{\text{S}_{1/2}, 1, 0} \nleftrightarrow \ket{\text{P}_{1/2}, 1, 0}$, this stochastically pumps the population into the qubit $\ket{0}$ state.
The stroboscopic pulsing prevents frequency mixing of the sidebands by the EOM which would lead to unwanted driving of the $1 \leftrightarrow 2$ transition and could thus excite population out of the $\ket{0}$ state.
This process typically achieves state preparation infidelities of $\sim$$8 \times 10^{-3}$, where the fidelity is limited by polarization impurities and the laser orientation with respect to the magnetic field~\cite{SuppMats}.
While one could attempt to improve these imperfections, it is technically difficult given vacuum window birefringence, finite quality retarders, and incomplete magnetic field control, all of which would be further exacerbated in a larger system with many spatially-separated qubits.

After polarization-limited state preparation, some population is left as error in the remaining Zeeman sublevels of $\text{S}_{1/2}$.
To flush out the leakage population in the $F=2$ manifold, we apply a weak pulse ($1$~$\mu$s, $90$~mW/cm$^2$) of 493~nm light carrying all polarizations and tuned to $(\text{S}_{1/2}, F=2) \rightarrow (\text{P}_{1/2}, F=2)$ to pump population equally into the three Zeeman sublevels of $(\text{S}_{1/2}, F=1)$.
During this pulse, $650$~nm light is also applied to pump population out of the $\text{D}_{3/2}$ manifold.
Next, to address the leakage population in $\ket{\text{S}_{1/2}, 1, \pm 1}$, we can use either microwave pulses to $\Ket{\text{S}_{1/2}, 2, \pm1}$ or shelving pulses to $\ket{\text{D}_{5/2},1,\mp1}$.
Under MAOP, two successive $\pi$-pulses $\Ket{\text{S}_{1/2},1,\pm 1} \rightarrow \Ket{\text{S}_{1/2},2,\pm1}$ ($\sim$$50~\mu$s) are applied to bring population into $F=2$ in preparation for another 493~nm flush pulse.
Each cycle of this procedure should ideally transfer the errors out of $F=1, m_F=\pm 1$ with microwaves, and then redistribute those errors equally back into the $F=1$ manifold with the $493$~nm flush pulse, leading to a $1/3$ reduction in error per cycle.
Since the microwave transitions are well separated from other transitions ($2\pi\times 3.45$~MHz) compared to our Rabi rates ($2\pi\times 10$~kHz), off-resonant excitations of the qubit state by the 493~nm enforce a more fundamental limit.

For NBOP, we apply two $\pi$-pulses of 1762~nm light $\Ket{\text{S}_{1/2},1,\pm 1} \rightarrow \Ket{\text{D}_{5/2},1,\mp1}$ ($57$, $46~\mu$s) to pump error into the $\text{D}_{5/2}$ state.
The orientation and polarization of the 1762~nm laser allows for only $\Delta m_F = \pm2$ transitions, meaning that any accidental excitation of the qubit state is suppressed by frequency separation, polarization, and laser $k$-vector.
Next, we apply a $4~\mu$s pulse of 614~nm light to address the dipole transition $(\text{D}_{5/2}, F=1) \rightarrow (\text{P}_{3/2}, F=0)$, from which population decays to the $F=1$ of the $\text{S}_{1/2}$ state with branching 73.8\% (zero branching to $F=2$)~\cite{dutta2016p32}.
To address any population that falls back into $\text{D}_{5/2}$, we apply sidebands to the 614~nm light that march the population to $(\text{P}_{3/2}, F=0)$: $2\rightarrow 1$, $3\rightarrow 2$, and $4\rightarrow 3$.
Similarly, 650~nm light is applied to repump population that falls into $\text{D}_{3/2}$.
With this scheme, only a small fraction of the population falls into $F=2$ states, reducing the need for the flush pulses.
We take advantage of this by running 35 cycles of NBOP where the final 5 cycles omit the flush pulse, lowering the error from  493~nm off-resonant excitations.



To characterize SPAM fidelity, we must also prepare $\ket{1}$ by applying a composite microwave $\pi$-pulse (CP Robust 180 sequence~\cite{Ba133, KnillPulse}) at the qubit frequency after preparing $\ket{0}$.
To distinguish between the two qubit states, we cabinet shelve population in $\ket{0}$ with three $\pi$-pulses to the states $\ket{\text{D}_{5/2}, 3, 2}, \ket{\text{D}_{5/2}, 2, -2}, \ket{\text{D}_{5/2}, 2, 2}$ (172, 55, 49~$\mu$s).
Due to magnetic field and laser intensity noise, these pulses are limited to fidelities of 0.960, 0.990, 0.988, respectively, but with three pulses we are able to shelve with an overall infidelity of $\left(4.7\pm0.2\right)\times10^{-6}$ assuming uncorrelated failure mechanisms.
After shelving, $493$~nm and $650$~nm light are applied with all sidebands to detect any population in $\text{S}_{1/2}$, $\text{P}_{1/2}$, and $\text{D}_{3/2}$.
Notably, this process only distinguishes population in $\ket{0}$ versus population in every other hyperfine state, including $\ket{1}$.

\begin{table}[!t]
    \centering
    \begin{tabular}{|c"c|c|}
				\hline 
        \multirow{2}{*}{Error Source} & \multicolumn{2}{c|}{Error ($\times10^{-5})$} \\
				\cline{2-3}
         & $\ket{0}$ State & $\ket{1}$ State\\
				\thickhline
         Depumping due to flush beam & $0.34$ & $0.34$ \\
				\hline
				 Depumping due to $1762$~nm beam & $<$$0.1$ & $<$$0.1$ \\
				\hline
         $\ket{0}\rightarrow\ket{1}$ transfer & --- & $2.97\pm0.33$ \\
				\hline
         Cabinet shelving infidelity & $0.47\pm0.02$ &  --- \\
				\hline
         $\text{D}_{5/2}$ decay & $1.50$ & --- \\
				\hline
				 Shelving through $\text{P}_{3/2}\rightarrow\text{D}_{5/2}$ & --- & $\leq$0.1 \\
				\hline
				 Histogram overlap error & $<$$0.1$ & $<$$0.1$ \\
				\hline
         Correlated errors (measured) & $2.8$ & $2.8$ \\
				\thickhline
         Subtotal (predicted) & $5.12$ & $6.21$ \\
				\hline
				 Total (predicted) & \multicolumn{2}{c|}{$5.67$} \\
				\thickhline
         Subtotal (raw data) & $8.79\pm1.83$ & $3.69\pm1.18$ \\
				\hline
         Total (raw data) & \multicolumn{2}{c|}{$6.15\pm1.08$} \\
				\thickhline
         Subtotal (corrected) & $8.79\pm1.83$ & $9.37\pm1.89$ \\
				\hline
         Total (corrected) & \multicolumn{2}{c|}{$8.99\pm1.31$} \\
				\hline
    \end{tabular}
    \caption{Error budget for 1762~nm narrow-band optical pumping SPAM results of Fig.~\ref{fig:histograms}(c). The final SPAM infidelities correct for the $\ket{0}$ state preparation error in the $\ket{1}$ error.}
    \label{table:error_budget}
\end{table}

Figure~\ref{fig:histograms}(a) presents the SPAM performance of both state preparation procedures over a variable number of cycles including the $493$~nm flush beam.
After $>$$25$ cycles of either MAOP or NBOP, we observe a drop in the SPAM infidelity of nearly two orders of magnitude compared to using polarization state preparation alone.
Both state preparation schemes initially exhibit close to the expected 1/3 reduction in error (gray line), but other error contributions begin to dominate at lower infidelities.
The performance of MAOP and NBOP appear comparable up until $30$ cycles with the flush pulse, but 1762~nm NBOP proves superior in a larger dataset ($10^6$ trials) at $35$~cycles, where we omit the $493$~flush pulse in the final five cycles of NBOP.
The fluorescence count histograms for this $35$~cycle data are shown in Fig.~\ref{fig:histograms}(b-c), and using the same discriminator (vertical dashed line) we extract SPAM infidelities of $\left(2.20 \pm 0.21 \right) \times 10^{-4}$ (MAOP) and $(9.0 \pm 1.3)\times10^{-5}$ (NBOP).


Because our measurement protocol only distinguishes between $\ket{0}$ (shelved to $\text{D}_{5/2}$ and observed as a lack of fluorescence) and not $\ket{0}$ (ion fluorescence), we can incorrectly identify the ion as being in the $\ket{1}$ state when it is actually in one of the other Zeeman levels of the $\text{S}_{1/2}$ manifold.
We correct for this by assuming any observed state preparation error for $\ket{0}$ also occurs for $\ket{1}$ (in addition to error in the microwave pulse required to prepare $\ket{1}$).
To apply this correction, we subtract known sources of measurement error from the $\ket{0}$ SPAM error, and add that to the observed $\ket{1}$ state SPAM error.
This correction is applied to all reported SPAM values in this work, and is described further in the Supplementary Materials~\cite{SuppMats}.

We list estimates for known error contributions to the higher fidelity NBOP result in Table~\ref{table:error_budget}, with extended discussion in the Supplementary Materials~\cite{SuppMats}.
We approximate the state preparation error after 30 flush cycles of NBOP to be equal to the MAOP value calculated from Eq.~\eqref{eq:eprep}, and further reduce this error by five no flush cycles (separately measured to reduce error by a factor of 3.2) to $3.4\times10^{-6}$.
%
%
%
Additional state preparation errors for the $\ket{1}$ state occur due to the infidelity of the composite microwave $\pi$-pulse, which we estimate to be $(2.97\pm0.33)\times10^{-5}$ by repeating up to 1000 consecutive composite pulses and extracting the average infidelity per pulse.
%
%
%
Measurement errors are dominated by shelved ions spontaneously decaying from the $\text{D}_{5/2}$ state into states bright to the measurement light, and these can be reduced by shortening our measurement time and increasing the shelving Rabi rates.
%
%
Other errors include correlated errors induced by collisions, laser instability, and magnetic field noise which we detail further in the Supplementary Materials~\cite{SuppMats}.
%

Our present implementation of these state preparation techniques takes about $3.5$~ms for 35 cycles using $\sim$50~$\mu$s microwave and 1762~nm $\pi$-times.
These long durations can result in hot ions, limiting achievable fidelities mainly through failed $\text{D}_{5/2}$ state shelving.
Moreover, this timescale is significantly longer than typical ion qubit gate times of $50~\mu$s, impacting overall circuit runtime.
We can reduce the required time for either scheme by performing both $\pi$-pulses in parallel.
Furthermore, if we allow an error of $10^{-5}$ due to off-resonant excitation of $\ket{0}$, then the minimum allowed microwave and $1762$~nm $\pi$-times are $12.5~\mu$s and $2.5~\mu$s, respectively, and could be lowered even further with pulse shaping methods.
We project that these improvements could cut down the state preparation time to $450~\mu$s (microwave) and $250~\mu$s (1762~nm).

In this work, we have achieved the highest reported SPAM fidelity of any qubit using a $^{137}\mbox{Ba}^+$ ion with nuclear spin $I=3/2$.
Our two similar state preparation techniques are generalizable to many ion species with higher nuclear spin $I>1/2$, facilitating quantum computation with new ion species with benefits like visible wavelengths and lighter masses.

\begin{acknowledgments}
The authors would like to thank Stephen Erickson, Chris Gilbreth, Colin Kennedy, Conrad Roman, and Jonathan Sedlacek for their suggestions and advice, as well as the rest of the Quantinuum team for their contributions.
\end{acknowledgments}


%

\end{document}


\title{High fidelity state preparation and measurement of ion hyperfine qubits with $I>\frac{1}{2}$}
\author{Fangzhao Alex An}
\email{Fangzhao.An@Quantinuum.com}
\affiliation{Quantinuum, Golden Valley, MN, USA}
\author{Anthony Ransford}
\email{Anthony.Ransford@Quantinuum.com}
\affiliation{Quantinuum, Broomfield, CO, USA}
\author{Andrew Schaffer}
\author{Lucas R. Sletten}
\affiliation{Quantinuum, Golden Valley, MN, USA}
\author{John Gaebler}
\affiliation{Quantinuum, Broomfield, CO, USA}
\author{James Hostetter}
\author{Grahame Vittorini}
\affiliation{Quantinuum, Golden Valley, MN, USA}

\date{\today}

\maketitle

\section{Energy level diagram of ${}^{137}\mathrm{Ba}^+$}

\begin{figure}[h!]
\centering
\includegraphics[width=0.45\linewidth]{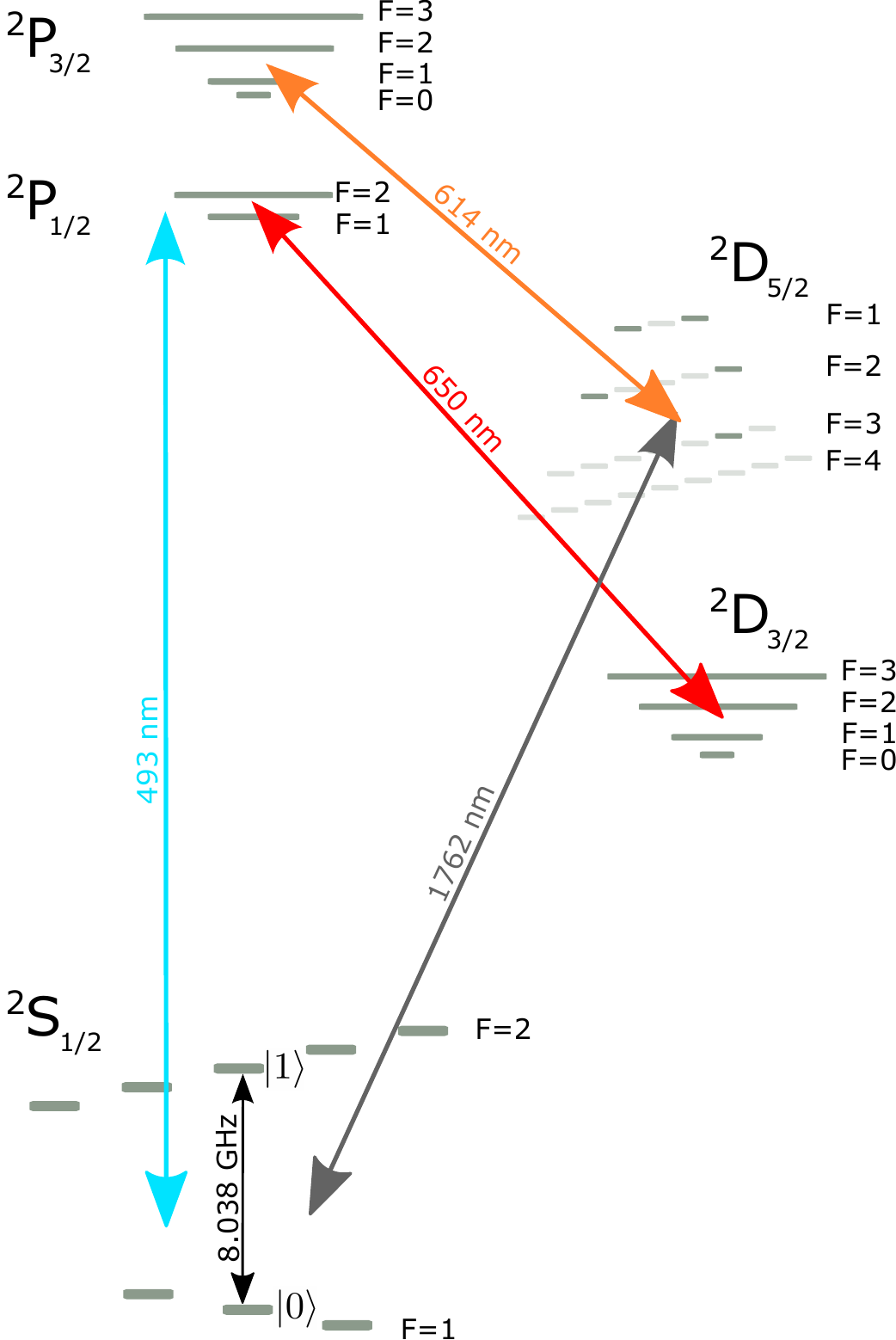}
\caption{\label{Level diagram}Level diagram of $^{137}\mathrm{Ba}^+$ highlighting the transitions addressed in this work: 493~nm cycling (aqua), 650~nm repump (red), 1762~nm shelving (gray), 614~nm deshelving (orange), and microwave transition between $^2\text{S}_{1/2}$ hyperfine levels (black).}
\label{fig:level_diagram_big}
\end{figure}

Figure~\ref{fig:level_diagram_big} shows the level structure of $^{137}\mathrm{Ba}^+$ and the transitions we address in this work.
We deal with the many hyperfine levels by applying frequency sidebands to all of our lasers using electro-optic modulators.
The relevant Zeeman levels in $^2\text{D}_{5/2}$ are highlighted: we shelve from $\ket{0}$ to $\ket{\text{D}_{5/2},2,\pm2}$ and $\ket{\text{D}_{5/2},3,2}$, and we use $\ket{\text{S}_{1/2},1,\pm1}\rightarrow\ket{\text{D}_{5/2},1,\mp1}$ for 1762~nm narrow-band optical pumping.

\clearpage

\begin{figure}[t!]
\centering
\includegraphics[width=0.5\linewidth]{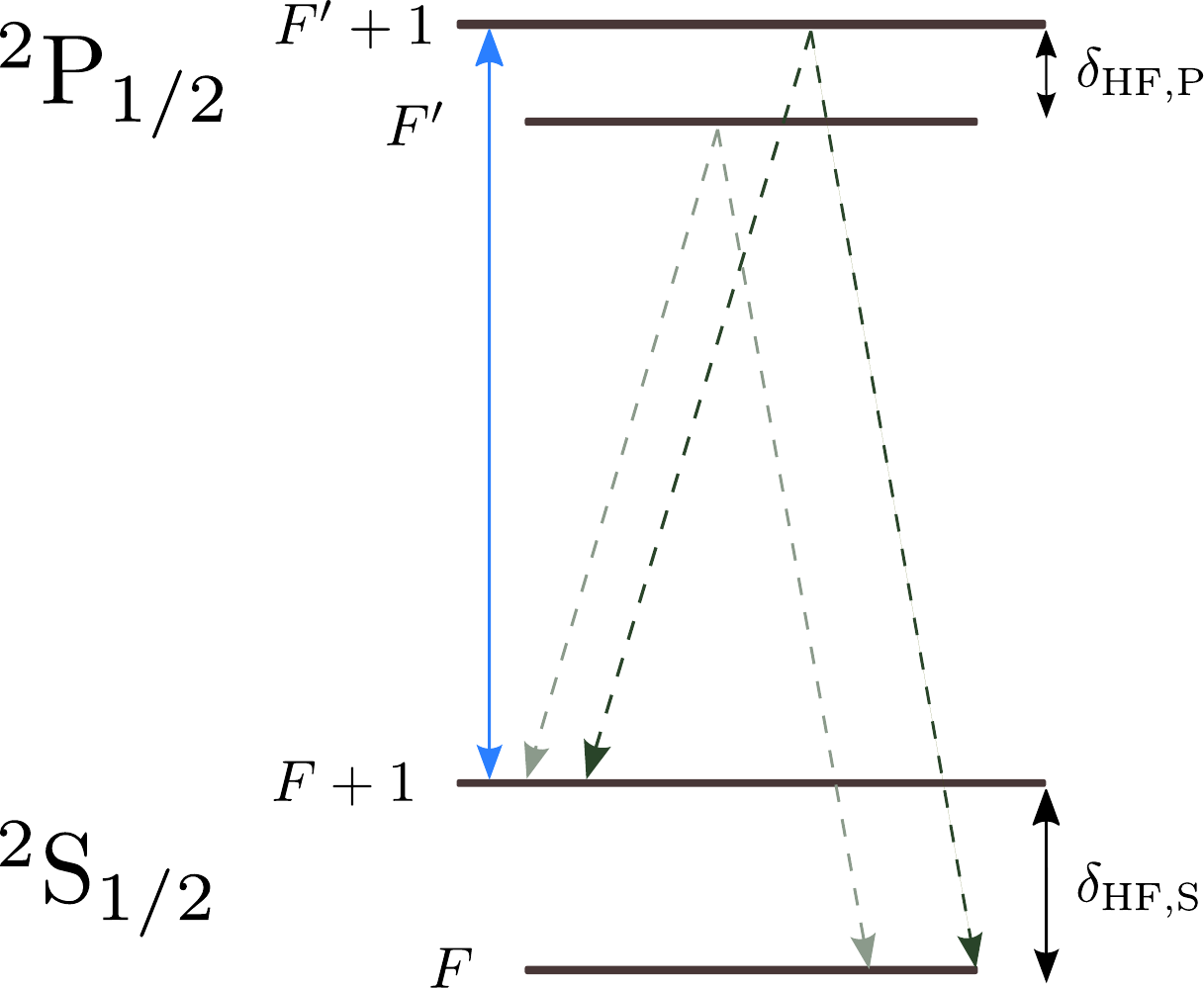}
\caption{\label{Theory Diagram}Level diagram of a typical alkaline earth ion with flush pulse (blue), flush branching (dark dashed lines) and error branching (light dashed lines).}
\label{fig:figure137}
\end{figure}

\section{Modeling Microwave-Assisted Optical Pumping State Preparation}
\label{sec:MAOP}

We consider a general alkaline earth ion with a level structure shown in Fig.~\ref{fig:figure137}.
Under our microwave state preparation technique, the flushing laser addresses the $F + 1 \rightarrow F' + 1$ transition, fully clearing out the $F + 1$ manifold.
Microwaves tuned near the hyperfine splitting $\delta_{\mathrm{HF},\mathrm{S}}$ then drive population from the unwanted Zeeman states in the $F$ manifold to the $F + 1$ manifold, and the cycle is repeated.
In the limit of low flush power and a large number of pulses, the steady state can be viewed as equal populations in the upper ($F+1$) and lower ($F$) manifold Zeeman states.
The time rate of change of the non-qubit Zeeman states is given by
\begin{equation}
\begin{aligned}
\frac{d\rho_{\neq |0\rangle}}{dt} = \eta_{_{F' + 1, F}} \frac{\gamma}{2}\rho_{\neq |0\rangle},
\end{aligned}
\label{eq:not_qubit}
\end{equation}

\noindent and the rate equation for the qubit state is given by

\begin{equation}
\begin{aligned}
\frac{d\rho_{|0\rangle}}{dt} = \Bigg[\eta_{_{F' + 1, F + 1}}\Big(\frac{\Gamma}{2\delta_{_{\mathrm{HF}, \mathrm{S}}}}\Big)^2 + \eta_{_{F', F + 1}}\Big(\frac{\Gamma}{2\delta_{-}}\Big)^2\Bigg]  \gamma\rho_{|0\rangle},
\end{aligned}
\label{eq:qubit}
\end{equation}

\noindent where $\eta_{i, j}$ denotes the branching from state $i$ to state $j$, $\gamma$ is the scattering rate, $\Gamma$ is the transition linewidth, and $\delta_{-} = \delta_{\mathrm{HF},\mathrm{S}}-\delta_{\mathrm{HF},\mathrm{P}}$ is the difference of the hyperfine splittings.
The factor of 2 in Eq.~\eqref{eq:not_qubit} comes from the steady state assumption of equal population in the two coupled Zeeman states.
Setting these rates equal to each other, we solve for the steady state and take the ratio of the populations as the state preparation error

\begin{equation}
\begin{aligned}
\epsilon_\mathrm{prep} = \frac{\Gamma^2}{2}\Bigg[\frac{1}{\delta_{_{\mathrm{HF}, \mathrm{S}}}^2} + \frac{\eta_{_{F', F + 1}}}{\eta_{_{F' + 1, F + 1}}}\frac{1}{\delta_{-}^2}\Bigg].
\label{eq:eprep}
\end{aligned}
\end{equation}

The first term is identical to twice the limit of state preparation in an $I=1/2$ ion and the second term is a correction factor that depends on the branching ratio and hyperfine splitting difference between the ground and excited states $\delta_{-}$.
The parameters for Eq.~\eqref{eq:eprep} and the associated fidelities for various $I>1/2$ hyperfine qubits are shown in Table~I of the main text.
We do not model the measurement errors in general because of the variety of high fidelity schemes available from coherent and incoherent shelving to quantum logic and quantum non-demolition~\cite{erickson2021high}.

\begin{figure}[h]
\centering
\includegraphics[width=\linewidth]{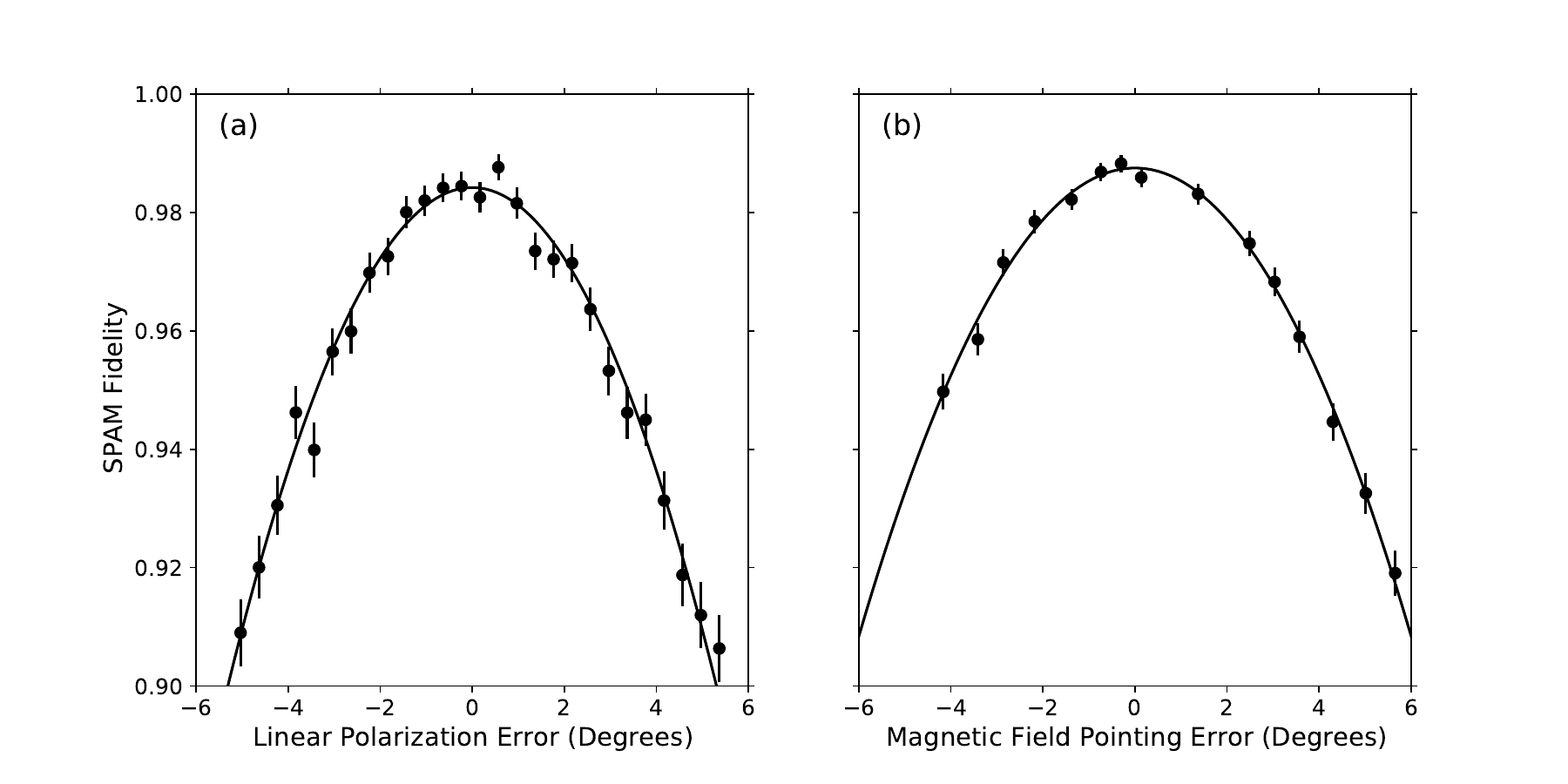}
\caption{\label{SvS} Errors in polarization SPAM can be exacerbated by \textbf{(a)} polarization errors in the optical pumping beam, shown here as a rotation of the linear polarization, or \textbf{(b)} incorrect alignment of the magnetic field to the \emph{k}-vector, demonstrated here using orthogonal magnetic field shims. Lines are added to guide the eye.}
\label{fig:pol_sp}
\end{figure}

\section{Polarization state preparation}

Using polarization state preparation alone, we can achieve SPAM infidelities of $(8.3\pm1.3)\times 10^{-3}$.
The fundamental limit is off-resonant excitation of $\ket{0}$, but we believe this fidelity is technically limited by polarization impurities and imperfect alignment of the the laser \emph{k}-vector orthogonal to the quantization axis.
To investigate these error mechanisms, we measured the infidelity as a function both magnetic field direction and polarization (Fig. \ref{fig:pol_sp}).
To vary the magnetic field direction, we measured SPAM fidelity while applying a shim field using coils orthogonally polarized to the quantization axis, recalibrating our shelving transitions to account for any changes to overall magnetic field amplitude.
To introduce polarization impurities, we simply rotated a half-wave plate before measuring SPAM fidelity.
While one could attempt to further improve these imperfections, it is technically difficult given vacuum window bi-refringence, finite quality retarders, and incomplete magnetic field control, all of which will be further complicated in a system with many spatially separated qubits.
Though we found optimal performance at a non-zero shim field, the coils were ultimately disconnected to reduce any magnetic field noise introduced to the shelving transitions and further SPAM improvements were achieved using the techniques described in the main text.

\section{Error Budget for 1762~nm narrow-band optical pumping SPAM}

Here we describe in more detail the errors in our measured SPAM infidelity using 1762~nm narrow-band optical pumping, listed in Table~II of the main text.

As described in Sec.~\ref{sec:MAOP}, in the limit of zero flush beam power and infinite time, MAOP reaches a steady state of equal error in the $F=1$ and $F=2$ manifolds of S$_{1/2}$, resulting in a $\ket{0}$ state preparation error $\sim$$1.1\times10^{-5}$ for $^{137}$Ba$^+$.
We assume that the 30 flush cycles of NBOP reaches a similar error.
To determine the effect of the final five cycles without flush pulses, we first prepare an ion in $\ket{0}$ to an infidelity of $7\times10^{-3}$ using the polarization-sensitive state preparation method only.
Then we apply five cycles of NBOP without the $F=2$ flush beam and find a reduction in the infidelity by a factor of 3.2 (versus 6.0 if the flush pulses are performed), which is likely limited by undesired decay to $F=2$ states.
%
Assuming the same error reduction rate at the end of the 30 flush cycles, we estimate our $\ket{0}$ state preparation error to be $(1.1\times 10^{-5}) / 3.2 = 3.4\times10^{-6}$.
%
%
%
%
We note that this state preparation error must be manually added to the $\ket{1}$ error, which we discuss below in Sec.~\ref{sec:sp_correction}.

Another possible state preparation error is off-resonantly driving the transition $\ket{0}\rightarrow\ket{\text{D}_{5/2},1,0}$ with the 1762~nm pulses during NBOP ($\Ket{\text{S}_{1/2},1,\pm 1} \rightarrow \Ket{\text{D}_{5/2},1,\mp1}$).
Due to the long lifetime of the D$_{5/2}$ state any off-resonant excitation should be infinitesimal, and the beam polarization and geometry also forbid this $\Delta m = 0$ transition, leading to an error less than $10^{-6}$.

The state preparation error of $\ket{1}$ additionally includes the infidelity of the microwave composite $\pi$-pulse~\cite{Ba133, KnillPulse} from $\ket{0}$ to $\ket{1}$.
We measure the infidelity of repeated composite $\pi$-pulses over a range of 0 to 1000 pulses, and perform a linear fit to extract an average infidelity of $(2.97 \pm 0.33)\times 10^{-5}$ per pulse.

%


%

\begin{figure}[b]
\centering
\includegraphics[width=\linewidth]{poison.pdf}
\caption{\label{poison} Poissonian fits to the dark (black) and bright (gold) SPAM histograms of Fig.~2(b-c) of the main text, with fitted average counts in legend. We choose our threshold (vertical dashed line) where the fits of the dark and bright intersect (16 counts). The fits do not overlap for $10^6$ counts, so we estimate $<$$10^{-6}$ overlap error.}
\label{fig:poison}
\end{figure}


Measurement errors are dominated by the infidelity of our cabinet shelving sequence and decay from the shelved state during measurement and shelving.
We measure the cabinet shelving fidelity by running $3\times10^5$ trials of NBOP $\ket{0}$ SPAM using only one shelving pulse at a time, and subtracting any errors from a simultaneous experiment using full cabinet shelving.
This way, we measure the fidelities of the shelving $\pi$-pulses to states $\ket{\text{D}_{5/2},3,2}$, $\ket{\text{D}_{5/2},2,-2}$, and $\ket{\text{D}_{5/2},2,2}$ to be $0.9598\pm0.0007$ ($\pi$-time 172~$\mu$s), $0.9904\pm0.0003$ (55~$\mu$s), and $0.9878\pm0.0004$ (49~$\mu$s), respectively.
Running the three shelving pulses in sequence achieves a total cabinet shelving error of $(4.67\pm0.24)\times10^{-6}$.
We treat this as a lower bound on the error, as this measurement does not account for possible slow magnetic field noise which may simultaneously reduce the fidelity of all three shelving pulses.

An ion shelved to D$_{5/2}$ may decay over the 350~$\mu$s measurement time as well as during cabinet shelving itself (172~$\mu$s, 55~$\mu$s, and 49~$\mu$s).
Given the 30.12~s lifetime of the $\text{D}_{5/2}$ state~\cite{DstateLifetime}, we can estimate the decay error considering only the measurement duration plus the final two shelving pulses (453~$\mu$s in total), giving $1.5\times10^{-5}$.
We also attempt to identify decays that actually occurred in the dataset by splitting the 350~$\mu$s measurement time into ten $35~\mu$s intervals.
In post-processing, we then track the counts over these intervals, and an accidentally bright error caused by a decay will register few counts in early segments before becoming bright in later ones.
%
We analyzed the measurements using a Bayesian update assuming Poissonian-distributed counts and pre-determined fluorescence rates; each measurement begins with a uniform prior that is updated for each successive segment based on the registered counts and the known distributions and ends once enough segments have been analyzed to cross a confidence threshold, here $10^{-6}$, or all ten segments have been considered.
Decays are then identified as shots that the threshold method determined to be bright but the Bayesian post-processing outcome was dark.
This method can only catch decays that occur after an integration time sufficient to pass the confidence threshold, which varies shot to shot but is 140~$\mu$s on average.
In the NBOP histogram data, we identify only one such decay during the remaining 210~$\mu$s which corresponds to roughly $2\times10^{-6}$ error during the 453~$\mu$s time the ion is shelved (measurement time plus second and third pulses of cabinet shelving).
In the MAOP data, we identify 7 decays, corresponding to $1.5\times10^{-5}$ decay error.
While we implemented this step in post-processing, the analysis could be implemented in real-time to conclude measurement once a given confidence threshold is reached, both shortening average measurement time and cutting down on errors from decays.


Additionally, this same technique allows us to monitor any bright counts that turn dark during the measurement duration.
Given a bright ion in $\text{S}_{1/2}$, the measurement beam at $493$~nm can off-resonantly drive the $\text{S}_{1/2} \leftrightarrow \text{P}_{3/2}$ transition at $455$~nm.
From $\text{P}_{3/2}$, the ion has a 26\% chance to decay into the dark state $\text{D}_{5/2}$.
%
%
%
This process should occur at less than $10^{-6}$, but we surprisingly see one such bright-to-dark error in post-processing the NBOP data.
%



Finally, measurement errors can also arise from overlap between the dark and bright state histograms.
We fit both MAOP and NBOP data to Poissonian distributions in Fig.~\ref{fig:poison}, observing no overlap between the fits for $10^6$ trials, and hence $<10^{-6}$ error.
We also use the overlap between the fits to determine the threshold value of $16$ counts (dashed vertical line).

A large error contribution comes from what we call ``correlated errors'' which appear as $\ket{0}$ errors occurring in bunches as large as 10-20 trials in a row, with fluorescence counts lower than what we would expect for a typical ``bright'' ion.
In taking SPAM data, we interleave trials preparing $\ket{0}$ and $\ket{1}$, so these correlated $\ket{0}$ errors also cause the following $\ket{1}$ counts to be erroneous, though they still appear bright (we account for this in the reported fidelities -- see Sec.~\ref{sec:sp_correction}).
We believe that these errors are caused primarily by background collisions with our ion, but also by other experimental imperfections such as a poor $493$~nm laser lock and noise on our trapping electrodes.
In the NBOP data set, we detect 28 such correlated bright errors, corresponding to a $2.8\times10^{-5}$ error contribution.


\section{Correction for $\ket{1}$ state preparation error}
\label{sec:sp_correction}
As described in the main text, our measurement technique can only distinguish $\ket{0}$ from all of the other Zeeman levels, so when we prepare $\ket{1}$ our measurement lumps together population in $\ket{1}$ along with all of the other non-$\ket{0}$ Zeeman levels of S$_{1/2}$.
More specifically, our state preparation schemes (MAOP or NBOP) prepare $\ket{0}$, leaving some errors in $\ket{\text{S}_{1/2},1,\pm1}$ and all five Zeeman levels of $(\text{S}_{1/2},F=2)$ (ignoring any small errors left in other states like D$_{3/2}$).
The microwave $\pi$-pulse to prepare $\ket{1}$ flips $\ket{0}$ with the state preparation error in $\ket{\text{S}_{1/2},2,0}$, and the subsequent cabinet shelving measures the state preparation error in only this one Zeeman level.
Thus to get a proper SPAM fidelity for $\ket{1}$, we need to account for any state preparation error left in the other six Zeeman levels.

We can estimate this error by looking at the error budget of Table~II of the main text.
We measure $(8.79\pm1.83)\times10^{-5}$ overall error for $\ket{0}$ SPAM (with 1762~nm NBOP), which is comprised of state preparation error and measurement error.
By subtracting the two known sources of measurement error from cabinet shelving and decay from the shelved state, we can set an upper bound on the $\ket{0}$ state preparation error.
As described above, we must only account for error in six of the seven Zeeman levels.
Assuming that the state preparation error is equally distributed among the Zeeman levels of S$_{1/2}$, we can make the correction
\begin{equation}
\epsilon_{1, \text{new}} = \epsilon_1 + \dfrac{6}{7} \left( \epsilon_0 - \epsilon_\text{decay} -\epsilon_\text{shelving} \right),
\label{eq:sp_correction}
\end{equation}
and we report this $\epsilon_{1,\text{new}}$ as the $\ket{1}$ infidelity $(9.37\pm1.89)\times10^{-5}$, and use it to calculate the total SPAM infidelity $(8.99\pm1.31)\times10^{-5}$.
This correction is applied to all reported SPAM fidelities in this paper.

\begin{figure}[h!]
\centering
\includegraphics[width=0.8\linewidth]{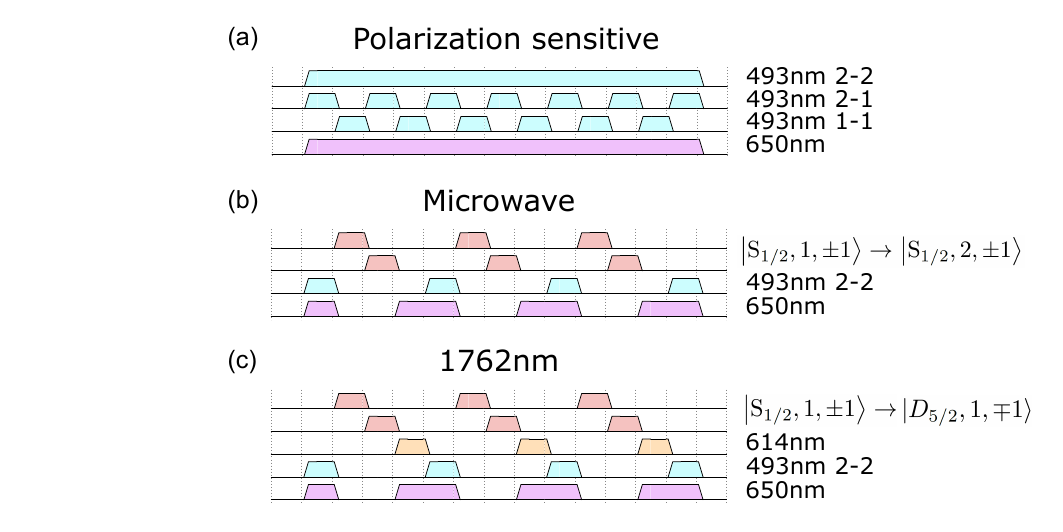}
\caption{Experimental pulse sequences for \textbf{(a)} polarization state preparation with stroboscopically pulsed 493~nm sidebands, \textbf{(b)} microwave-assisted optical pumping, and \textbf{(c)} 1762~nm narrow-band optical pumping including the $493$~nm flush beam. Axes are not to scale.
}
\label{fig:timing_diagram}
\end{figure}

\section{State preparation pulse sequences}
Figure~\ref{fig:timing_diagram} shows the pulse sequences we use for state preparation.
Under polarization state preparation, $\pi$-polarized 493~nm light is applied to the ion to pump out all Zeeman states of $\text{S}_{1/2}$ except the qubit state $\ket{0}$ by taking advantage of the forbidden dipole selection rule $\ket{L,F,m_F=0}\nleftrightarrow\ket{L+1,F,m_F=0}$.
The carrier at $F=2 \leftrightarrow F=2$ (2-2) pumps out $\ket{\text{S}_{1/2},F=2,m_F\neq0}$, while sidebands at 2-1 and 1-1 pump out $\ket{\text{S}_{1/2},F=2,m_F\neq\pm2}$ and $\ket{\text{S}_{1/2},F=1,m_F\neq0}$, respectively.
The sidebands must be pulsed stroboscopically to avoid mixing at the EOM, as the beat note drives the unwanted 1-2 transition.
650~nm light is applied throughout to repump any ions that fall into $\text{D}_{3/2}$.

In microwave-assisted optical pumping, $493$~nm light with all polarizations flushes out the $(\text{S}_{1/2},F=2)$ manifold, and $650$~nm light is simultaneously applied to repump any ions that fall into $\text{D}_{3/2}$.
Then, microwaves $\pi$-pulses move errors from $F=1$ up to $F=2$ by addressing $\ket{\text{S}_{1/2},1,\pm1}\rightarrow \ket{\text{S}_{1/2},2,\pm1}$, and these errors are then redistributed to $F=1$ with the next flush pulse.

In 1762~nm narrow-band optical pumping, a similar $493$~nm flush pulse clears out the $F=2$ errors, but the $F=1$ errors are moved to $\text{D}_{5/2}$ with $1762$~nm $\pi$-pulses $\ket{\text{S}_{1/2},1,\pm1}\rightarrow \ket{\text{D}_{5/2},1,\mp1}$ ($\Delta m_F = \pm2$).
These errors are then deshelved with the $614$~nm laser to $(\text{P}_{3/2}, F=0)$, from which they predominantly decay back into $(\text{S}_{1/2}, F=1)$.

%
